\documentclass[sn-mathphys-num]{sn-jnl}%
\pdfoutput=1
\usepackage{graphicx}%
\usepackage{multirow}%
\usepackage{amsmath,amssymb,amsfonts}%
\usepackage{amsthm}%
\usepackage{mathrsfs}%
\usepackage[title]{appendix}%
\usepackage{xcolor}%
\usepackage{textcomp}%
\usepackage{manyfoot}%
\usepackage{booktabs}%
\usepackage{algorithm}%
\usepackage{algorithmicx}%
\usepackage{algpseudocode}%
\usepackage{listings}%

\theoremstyle{thmstyleone}%

\theoremstyle{thmstyletwo}%

\theoremstyle{thmstylethree}%

\newcommand{\todo}[1]{\textcolor{red}{TODO: #1}}

\renewcommand{\todo}[1]{}

\newcommand{\modelname}{NeCS}
\newcommand{\norm}[1]{\left\lVert#1\right\rVert}

\begin{document}

\title[Article Title]{Understanding complex crowd dynamics with generative neural simulators}

\author[1]{\fnm{Koen} \sur{Minartz}}\email{k.minartz@tue.nl}

\author[2,4]{\fnm{Fleur} \sur{Hendriks}}\email{f.hendriks@tue.nl}

\author[1]{\fnm{Simon Martinus} \sur{Koop}}\email{s.m.koop@tue.nl}

\author*[3,4]{\fnm{Alessandro} \sur{Corbetta}}\email{a.corbetta@tue.nl}

\author*[1,4]{\fnm{Vlado} \sur{Menkovski}}\email{v.menkovski@tue.nl}

\affil[1]{\orgdiv{Department of Mathematics and Computer Science}, %

\orgname{Eindhoven University of Technology}%
}

\affil[2]{\orgdiv{Department of Mechanical Engineering},%

\orgname{Eindhoven University of Technology}%
}

\affil[3]{\orgdiv{Department of Applied Physics and Science Education}, %

\orgname{Eindhoven University of Technology}%
}

\affil[4]{\orgdiv{Eindhoven Artificial Intelligence Systems Institute}%
}


\abstract{

Understanding the dynamics of pedestrian crowds is an outstanding challenge crucial for designing efficient urban infrastructure and ensuring safe crowd management. To this end, both small-scale laboratory and large-scale real-world measurements have been used. However, these approaches respectively lack statistical resolution and parametric controllability, both essential to discovering physical relationships underlying the complex stochastic dynamics of crowds.
Here, we establish an investigation paradigm that offers laboratory-like controllability, while ensuring the statistical resolution of large-scale real-world datasets. Using our data-driven Neural Crowd Simulator (\modelname{}), which we train on large-scale data and validate against key statistical features of crowd dynamics, we show that we can perform effective surrogate crowd dynamics experiments without training on specific scenarios. We not only reproduce known experimental results on pairwise avoidance, but also uncover the vision-guided and topological nature of N-body interactions. These findings show how virtual experiments based on neural simulation enable data-driven scientific discovery.    

\todo{
submit}

\todo{requirements:}
\todo{abstract:up to 150 words}
\todo{total article $\leq$ 3500 words excl abstract, methods, references, fig legends}
\todo{up to 6 display items (figures and/or tables}
\todo{Introduction, Results, Discussion, Methods structure}
\todo{discussion does not have subheadings}
\todo{guideline 50 references}}

\maketitle

\section{Introduction}\label{sec:main}

Establishing a physical understanding of pedestrian crowd dynamics is not only a fundamental challenge in active matter physics~\cite{corbetta-annurev-2023}, but also key in societal applications, ranging from infrastructural design to large-scale crowd management~\cite{still2021applied, feliciani2023trends}. 
From a physics perspective, crowds constitute an active matter system~\cite{marchetti2013hydrodynamics} with nonlinear $N$-body interactions~\cite{araujo2023steering} in which physical and psychological factors interplay at different spatiotemporal scales~\cite{hoogendoorn2004pedestrian}.
Unsurprisingly, crowd dynamics show strong fluctuations from statistical averages, even in terms of velocities or distances~\cite{corbetta-pre-2018,pouw-trc-2024,zanlungo2014potential}. In aggregate, however, trajectories exhibit distinct and reproducible statistics~\cite{corbetta-pre-2017,zanlungo2014potential}. %
Models considering pedestrian dynamics at the individual level must therefore reflect this probabilistic nature.
Treating pedestrians as inertial agents~\cite{cristiani-book-2014}, this entails modeling the time-dependent 
probability $p$ of the instantaneous acceleration of each individual: %
\begin{equation}\label{eq:first-probab-accel}
\ddot{x}_i(t)  \sim p_\theta \left(\ \cdot\ \mid \,\{x_j(\tau),\,  \tau \leq t,\, j=1,\ldots,N \} \, \right)\qquad i = 1,\ldots, N , %
\end{equation}
where $x_i = x_i(t)$ is the spatial position at time $t$ of the $i$-th individual in a crowd of $N$, $\theta$ denotes the model parameters, and the conditioning variables emphasize dependencies on neighbors and past trajectories.

Over the years, the \emph{social force} paradigm~\cite{helbing-pre-1995} has become one of the most diffused yet criticized~\cite{chen2018social,Haghani_Ronchi_2024} models for Eq.~\ref{eq:first-probab-accel}.  
It adopts a Langevin-like perspective where individual accelerations are governed by fluctuating forces:%
\begin{align}\label{eq:social-force-linear}
    \ddot{x}_i 
    &= f(x_i, \dot{x}_i, \theta_i) + \sum_{i \neq j} g(x_j - x_i, \theta_{ij}) + \epsilon \text{,}
\end{align}
where $f$ represents individual propulsion in dependence of position, velocity, and pedestrian-specific parameters $\theta_i$,
$g$ models the interaction between two pedestrians, modulated by their relative position and interaction-specific parameters, $\theta_{ij}$, and a $\delta$-correlated white noise $\epsilon$ mimics stochastic fluctuations. 
Eq.~\ref{eq:social-force-linear} shows remarkable statistical agreement with experimental data for prototypical scenarios like individual fluctuations, pairwise avoidance or dynamics of small social groups~\cite{corbetta-pre-2017, corbetta-pre-2018, zanlungo2014potential}, and qualitatively explains emergent phenomena like lane formation, intermittency, and jamming~\cite{helbing-pre-1995,bottinelli2016emergent}. 
Nevertheless, our limited understanding of crowd dynamics is reflected by oversimplifying assumptions like pairwise additivity of the interactions and linearly superimposed stochasticity, which are certainly false for more general and complex settings~\cite{corbetta-annurev-2023}.

To improve our understanding, the community has relied on either  
laboratory experiments~\cite{boltes2013collecting, Haghani2020}, or on large-scale real-world measurements~\cite{Pouw_Willems_van_Schadewijk_Thurau_Toschi_Corbetta_2022, corbetta-pre-2018, pouw2024arxiv, pouw2024dataset}.
Although laboratory experiments offer full control over experimental scenarios, they hardly produce sufficient data to characterize stochastic behavior due to unavoidably limited repetitions~\cite{corbetta-annurev-2023}. In contrast, real-world measurements enable large-scale data collection, but yield observations consisting of multiple intertwined scenarios that appear randomly~\cite{corbetta-pre-2018} and where all physical processes at play are entangled. 
Additionally, despite their scale, real-world measurements still provide only a sparse coverage of the virtually infinite set of scenarios that are admissible in reality. 
As a result, isolating physical effects from real world data has required labor-intensive, tailor-made analyses even for prototypical cases~\cite{corbetta-pre-2018,zanlungo2014potential}, and is likely infeasible for more complex, yet crucial, $N$-body interaction scenarios.
Consequently, efficiently extracting physical interaction relationships from large-scale measurements remains an outstanding challenge, and a central bottleneck towards an improved understanding of Eq.~\ref{eq:first-probab-accel}.

\begin{figure}[t]
    \centering
    \includegraphics[width=\linewidth]{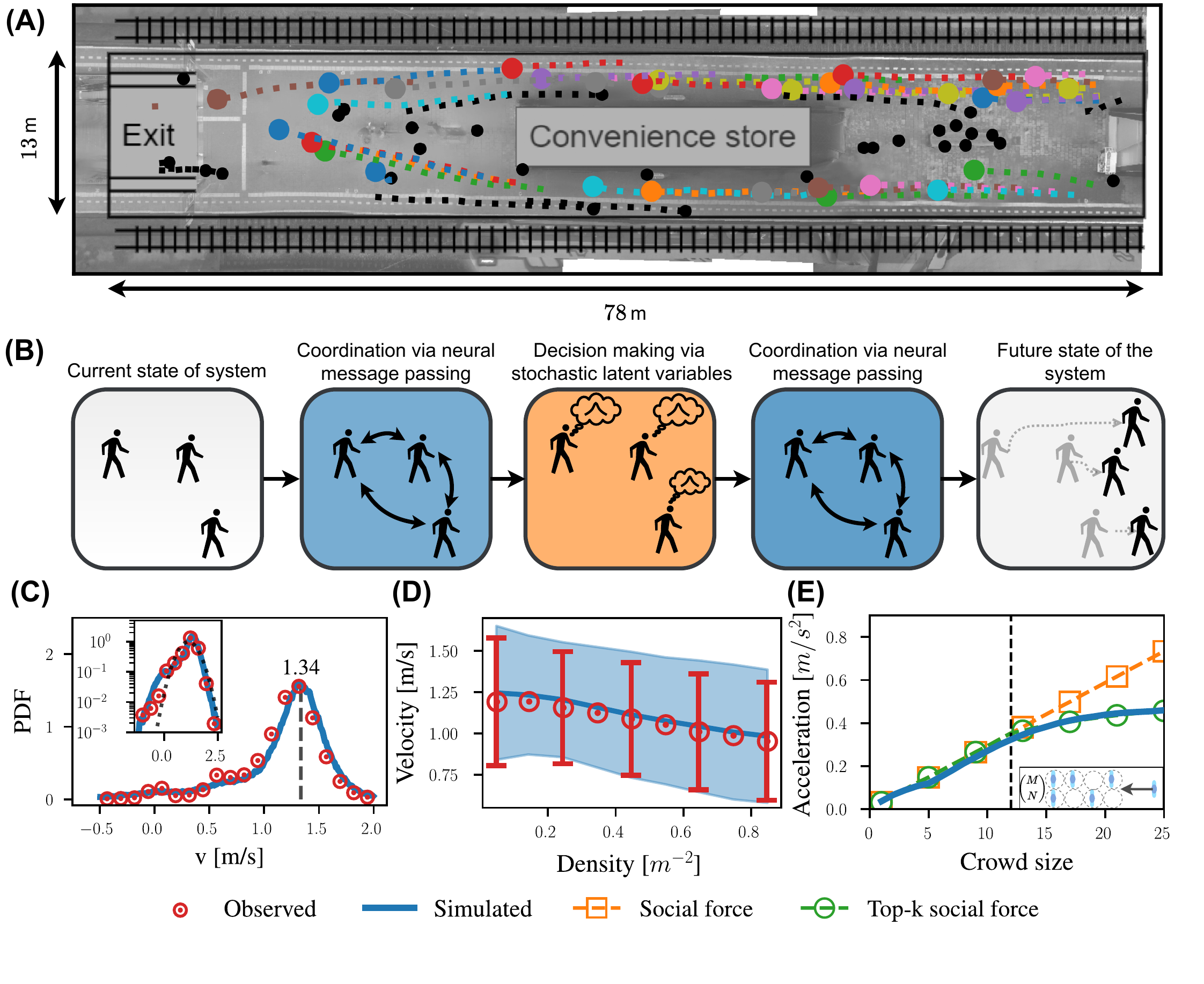}
    \caption{\textbf{(A)} Snapshot of a model simulation. Colored circles indicate positions of simulated pedestrians, black circles indicate pedestrian trajectories upon which the simulation is conditioned, dotted tails indicate a 10 second history out of the total 60 second simulation horizon. \textbf{(B)} Sketch of modeling approach. The model uses a message passing graph neural network to construct a representation of the current state. Randomness is modeled with latent variables at the individual level, conditioned on the constructed representation. The representation and latent variables are decoded to predict the next state. \textbf{(C)} Distribution over pedestrians' horizontal velocity component of both observations and simulated data. The black dotted line in the inset PDF plot indicates a fitted Gaussian distribution. \textbf{(D)} Fundamental diagram showing how pedestrian velocity decreases as density increases, for both simulated and observed data. The solid lines and markers indicate the mean velocity, while the shaded area and error bars indicate the standard deviation. \textbf{(E)} Virtual experiment measuring the social force on a pedestrian walking towards a crowd of $N$ static pedestrians. The plot shows the mean net force of the crowd on the pedestrian of the standard social force model, its top-$k$ variant, and of~\modelname{}. The vertical dashed line indicates $N^* \approx 12$. The inset shows a sketch of the surrogate experiment setup (see Figure~\ref{fig:model-interpretation} for more details).}
    \label{fig:intro-figure}
\end{figure}

Recently, Machine Learning (ML), and geometric deep learning in particular, has demonstrated great potential for data-driven modeling of complex dynamical systems~\cite{Battaglia2016interaction, sanchez-gonzalez2020, li2021fno, Li2024, Azizzadenesheli2024}. 
To evaluate the quality of such models, point-wise metrics such as mean squared error remain the dominant approach, including in the field of pedestrian trajectory modeling~\cite{Alahi2016sociallstm, Mohamed2022socialimplicit, Korbmacher2022, Mangalam2020PECnet, Gupta_2018_socialgan, Amirian_2019_socialways, Salzmann2020trajectron++, Wang2022SGNet, Yuan_2021_Agentformer, Mangalam_2021_YNet, tang2021socialpostcollapse}.
However, to trust the model's predictive abilities, let alone use it to extract any physical understanding, it is essential to validate against problem-specific metrics. For stochastic dynamics, this mandates going beyond point-wise errors and comparing the complete statistical portrait of the dynamics.

In this work, we propose \emph{virtual surrogate experiments} as an investigation paradigm, which we employ to investigate interactions in crowds. To this end, we introduce \modelname{}, short for Neural Crowd Simulator, illustrated in Figure~\ref{fig:intro-figure}. \modelname{} acts as a surrogate model for the extensive experiments required to characterize crowd dynamics in heterogeneous conditions. These surrogate experiments provide both the required parametric control that is infeasible in real-world settings and the necessary scale that is unattainable with laboratory experiments.
We first validate our approach by reproducing previously obtained results on pairwise collision avoidance~\cite{corbetta-pre-2018}. \modelname{} shows great agreement with measured pairwise
avoidance dynamics, and the social force field predicted by \modelname{} aligns remarkably well with~\cite{corbetta-pre-2018}. Building on this validation, we study more complex $N$-body interactions, for which to our best knowledge no prior experimental studies with robust statistical support exist. We find that the repulsive force on an individual facing a crowd scales like a vision-constrained top-$k$ social force model, which considers only the $k$ strongest forces ($k \approx 12$, Fig.~\ref{fig:intro-figure}E), as opposed to the linear scaling of Eq.~\ref{eq:social-force-linear}.

Our approach hinges on the combination of an architecture
at the interface of graph neural networks (GNNs)~\cite{Gilmer2017mpgnn}, recurrent neural networks~\cite{Hochreiter1997LSTM}, and conditional variational autoencoders~\cite{Kingma2014vae, Sohn2015cvae} and training data consisting of large-scale real-world measurements~\cite{pouw2024arxiv, pouw2024dataset}. We rigorously validate \modelname{} and show that it achieves a great level of statistical agreement with the observed dynamics (e.g., Figures~\ref{fig:intro-figure}C and~\ref{fig:intro-figure}D). 
Finally, key for reliable virtual experiments is for the analyzed scenarios to be in-distribution with respect to our training data (yet possibly sparsely occurring): a necessary condition for model interpolation, which we ensure in our analysis.

This paper is structured as follows: in Section~\ref{sec:problem}, we introduce the data used for model training and validation. Section~\ref{sec:results} explains the formulation of the generative model, and discusses its statistical validation. In Section~\ref{sec:physical_interp}, we report on our surrogate experiments. The discussion in Section~\ref{sec:conclusion} concludes the paper.

\section{Measurements and data}\label{sec:problem}

We consider high-fidelity crowd tracking measurements collected in April and May 2022 on a platform at Eindhoven Centraal train station (the Netherlands). These crowd dynamics, driven by a repeating train timetable, are recurrent and reproducible, and have been recorded via overhead sensors at $10\,$Hz temporal resolution~\cite{pouw2024arxiv, pouw2024dataset}.

We specifically focus on the flow of pedestrians moving towards the platform exit, depicted on the left side of Fig.~\ref{fig:intro-figure}A. These individuals, typically alighting from trains, maintain a sustained velocity over large spatial scales in short timeframes. This allows us to analyze consistent dynamics while capturing the complex interaction physics at various densities, from diluted conditions up to 1 ped/m$^2$. 

We preprocess the tracking data in~\cite{pouw2024dataset} by segmenting it into one-minute intervals. We keep only segments with: I) at least $10$ and at most $100$ distinct pedestrians; II) an average velocity  at least $0.25\,m/s$ in the direction of the exit; and III) at least $70\%$ of the individuals moving towards the exit. Pedestrians moving at least $15\,$m towards the exit are classified as part of the flow in that direction, and are the target of \modelname{}'s simulations. The remaining pedestrians serve as fixed inputs to condition Eq.~\ref{eq:first-probab-accel}, enabling us to study their influence on pedestrians moving towards the exit.

Overall, this procedure yields $55$ hours of measurements, with between $0$ and $90$ people present at any given time in a segment. We employ 80\% of the data for training, 10\% for validation, and 10\% for testing. Finally, we consider a 1:10 temporal subsampling
to remove small-scale fluctuations such as swaying~\cite{willems-scirep-2020}, while retaining all key features such as avoidance.

\section{Model formulation and statistical validation}\label{sec:results}

\paragraph{Model formulation}\label{sec:ml-formulation}
To approximate the distribution over the accelerations $\ddot{x}_i(t)$ of each pedestrian $i$ (Eq.~\ref{eq:first-probab-accel}), we condition the model on the current state as well as on the \emph{history}
of past states: 
\begin{alignat}{10}\label{eq:dist_acc_app}
    p_\theta \left(\ddot{x}_i(t) \mid s(t), h(t) \right)  &= p_\theta \left(\ddot{x}_i(t) \mid s_i(t), h_i(t),\ \left\{ s_j(t), h_j(t) \mid j \in \mathcal{N}(i) \right\} \right),\\
    s_i(t) &= [x_i(t), \dot{x}_i(t)] &&\left(\text{state}\right), \notag\\
    h_i(t) &= \left\{s_i(\tau) \mid \tau < t \right\} &&\left(\text{history}\right),\notag\\
    \mathcal{N}(i) &= \{j \mid \text{$j$ and $i$ are neighbors}\} &&\left(\text{neighborhood}\right).\notag
\end{alignat} 
In words, we consider $p_\theta$ depending on the state and history of pedestrian $i$ as well as their neighbors.
Eq.~\ref{eq:dist_acc_app} is parameterized with a conditional Variational Autoencoder~\cite{Kingma2014vae, Sohn2015cvae}, which autoregressively defines a probabilistic model over full trajectories. After training, we explore the trade-off between trajectory diversity and fidelity via a \emph{temperature} parameter $T$, analogous to autoregressive large language models~\cite{Vaswani2017Transformer}. $T$ scales the standard deviation of the model's decoder during sampling. We consider two variants with temperature values $T=T^*>0$ ($T^* = 0.1$) and $T=T^0=0$, respectively. Further details on the architecture, generative modeling approach and simulation procedure are in Section~\ref{sec:methods}.

\paragraph{Statistical model validation}
We evaluate model accuracy along three components of crowd dynamics: (1) ensemble statistics, assessing the model's capability to reproduce probability distributions over individual features; (2) interaction structure, capturing the social interactions in crowds; and (3) Lagrangian properties, characterizing pedestrian behavior over time. These components collectively ensure accuracy along fundamental physical aspects of crowds: probabilistic dynamics (1) with N-body interactions (2) that evolve the system over time (3).

\begin{figure}
    \centering
    \includegraphics[width=\linewidth]{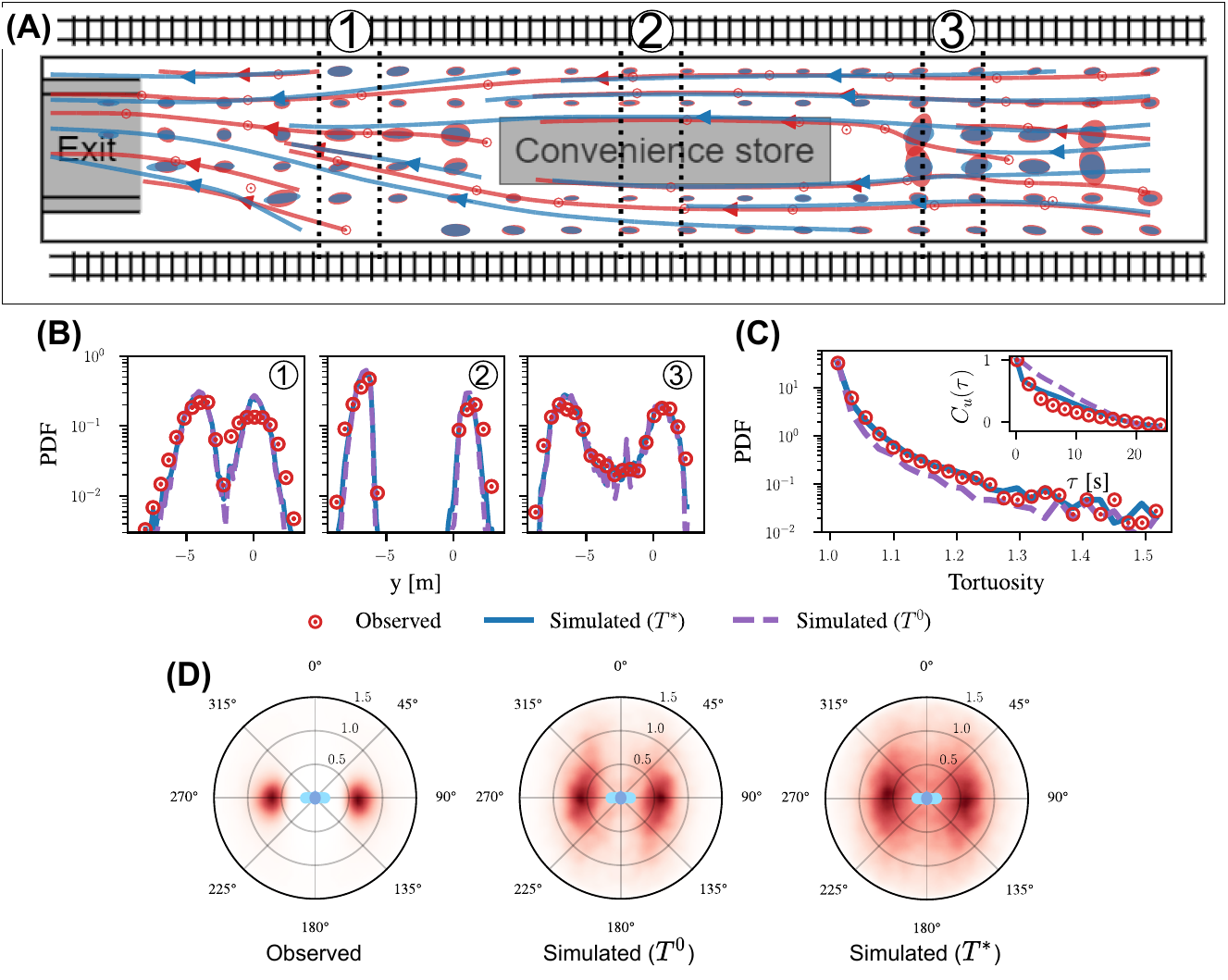}
    \caption{Model validation. \textbf{(A)} Mean velocity field $\bar{v}(x, y)$ at various locations $(x, y)$ in the station, visualized as flow lines along the field $\bar{v}$. The ellipsoids indicate the covariance matrix of the velocity vector at each location. \textbf{(B)} PDFs over simulated and observed vertical positions for various slices on the map in (A), indicated by the numbers \textcircled{1}-\textcircled{3}. \textbf{(C)} PDF of the tortuosity of observed trajectories and trajectories simulated by the two model variants. The inset plot shows the autocorrelation function of the vertical velocity component. \textbf{(D)} Density of the relative position of the relative position of the nearest neighbor for observed and simulated data. For $T=0$, focusing on sample fidelity, the distribution of the relative position is more focused than for $T=0.1$.} 
    \label{fig:model-validation}
\end{figure}

\vspace{0.125cm}
\noindent\textbf{Ensemble statistics.} %
Figure~\ref{fig:intro-figure}C shows the distribution of the longitudinal velocity component. \modelname{} replicates both the Gaussian-like structure around the most frequent walking velocity ($|\langle v \rangle| = 1.34\,$m/s) as well as the non-Gaussian tails. %
Additionally, Figure~\ref{fig:model-validation}A shows a streamplot illustrating the average pedestrian velocity at various locations, superimposed with covariance ellipses demonstrating directional variability. \modelname{} simulations align well with the mean velocity field, and the covariance shows a similar structure and scale for most locations, confirming accuracy in capturing position-dependent dynamics.
Figure~\ref{fig:model-validation}B depicts the y-coordinate distributions for longitudinal segments as marked in Figure~\ref{fig:model-validation}A. The bimodal structure with rapidly decaying densities around the center arises due to the presence of the convenience store. \modelname{} captures this decay accurately, indicating that model imperfections do not propagate to unrealistic positional distributions over the minute-long simulation horizon.

\vspace{0.125cm}
\noindent\textbf{Temporal structure.}
We characterize the temporal structure through tortuosity and autocorrelation analyses. Tortuosity is the ratio between actual and shortest possible trajectory length, and has previously been used to characterize insect flying paths~\cite{Jn2020tort}. Figure~\ref{fig:model-validation}C shows that the tortuosity distribution follows a sharp exponential decay between $1$ and $1.05$, covering trajectories close to the shortest possible path, followed by a milder exponential decay until tortuosity $1.5$. Both decay rates are reproduced by~\modelname{}: $T=T^0$ generally leads to marginally lower tortuosities than in the measurements, while $T=T^*$ aligns with the observations. 
 
The inset in Figure~\ref{fig:model-validation}C reports the autocorrelation of the transversal velocity component. Correlation functions have been routinely used to characterize Langevin-like time processes in general and pedestrian dynamics specifically~\cite{corbetta-pre-2017}. The autocorrelation decays at a similar rate as the observations when $T=T^*$, while $T=T^0$ produces a milder decay. Thus, $T=T^*$ results in more stochastic fluctuations in trajectories compared to $T=T^0$. We retain the case $T=T^*$ unless otherwise mentioned.

\vspace{0.125cm}
\noindent\textbf{Interaction.} %
We resort here to two statistical indicators of interactions: the fundamental diagram and nearest-neighbor distributions.
The fundamental diagram~\cite{Vanumu2017} represents how walking speed decreases as density increases, due to mutual interactions and velocity fluctuations~\cite{corbetta-annurev-2023}.
Figure~\ref{fig:intro-figure}D reports the probabilistic counterpart of the fundamental diagram to reflect the presence of fluctuations~\cite{pouw-trc-2024}. 
The mean velocity decays at the same rate in the simulations and in the measurements, and the spread in velocity values of the simulations and observations agree for all density values.

Figure~\ref{fig:model-validation}D depicts the relative positions of the nearest neighbors. As previously reported~\cite{Porzycki2014, pouw-trc-2024}, this position follows a bimodal distribution, with peaks at lateral distances of $0.5$-$0.75\,$m. 
For both $T^0$ and $T^*$,~\modelname{} produces a distribution with two modes at the right relative position. $T=T^*$ produces densities that are less concentrated than $T^0$, suggesting that $T=T^0$ enables the model to produce more strongly peaked distributions, whereas $T=T^*$ leads to higher variance as expected from the increased stochastic fluctuation (Figure~\ref{fig:model-validation}C). This illustrates a trade-off between sample fidelity ($T^0$) and diversity ($T^*$), which is typical in generative modeling~\cite{xiao2022trilemma}.

\section{Understanding pedestrian interactions with virtual surrogate experiments}\label{sec:physical_interp}
We  leverage \modelname{} to simulate prototypical scenarios in which we can investigate the structure of $N$-body interactions in pedestrian crowds. These simulations serve as virtual surrogate experiments and provide the same controllability of laboratory conditions, while allowing for virtually infinite repeated trials. 
Such surrogate experiments are meaningful only if the investigated scenarios are in-distribution, that is, if similar scenarios are present in the training data. We therefore accompany each experiment with plots reporting frequencies with which such similar cases are observed in the training set.

\begin{figure}
    \centering
    \includegraphics[width=\linewidth]{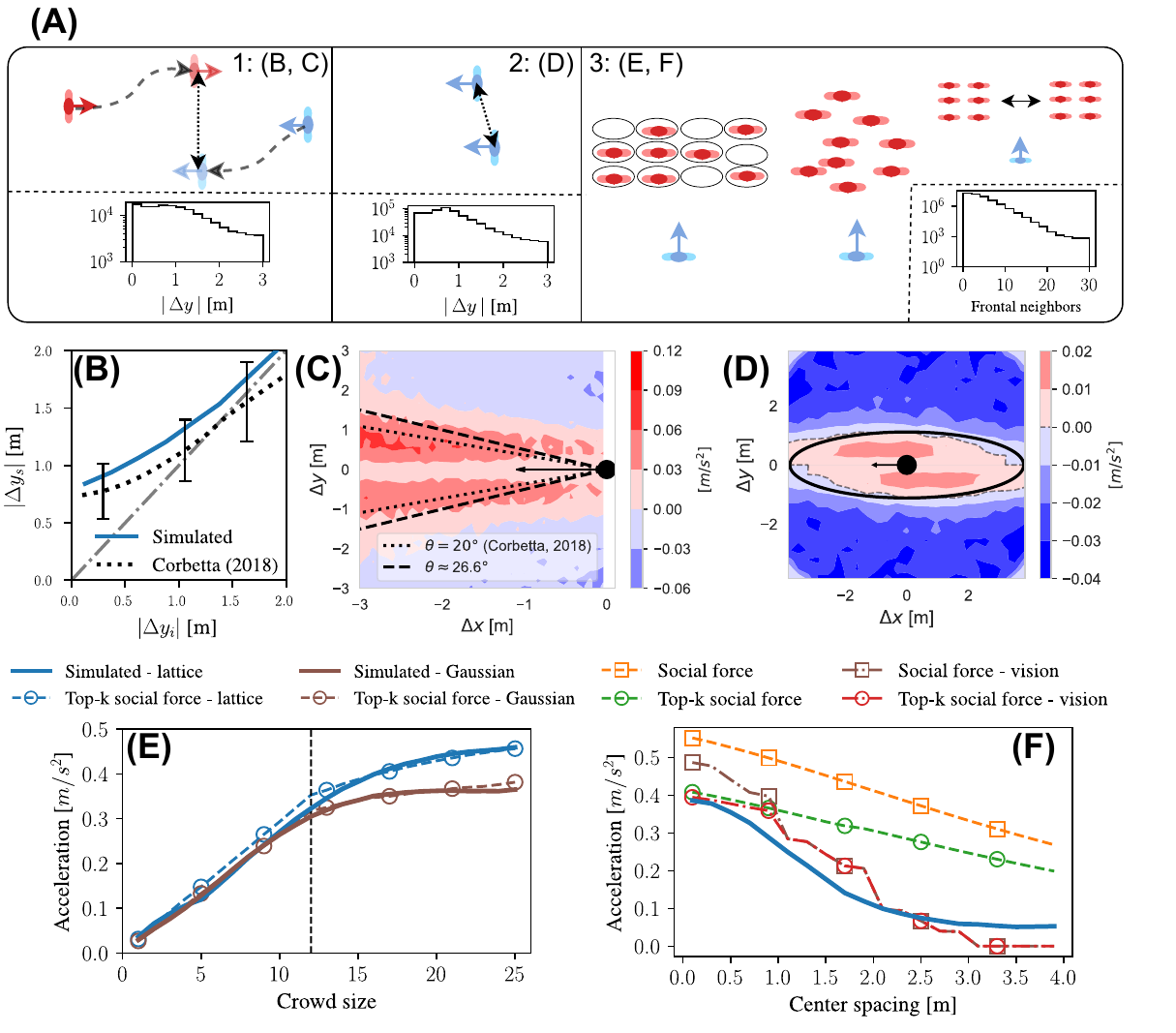}
    \caption{Understanding crowd physics by using \modelname{} for virtual surrogate experiments. \textbf{(A)} Sketches of experiments considered in this paper. We investigate the avoidance of pairs of pedestrians walking in opposite directions, the interactions of pedestrians walking in the same direction, and the strength of the force depending on the size and configuration of a static crowd in front of a pedestrian. The training data is analyzed to establish coverage of similar scenarios for each experiment, shown in the histograms. 
    \textbf{(B)} Counterflow transfer function, depicting the lateral distance at the moment of passing $|\Delta y_s|$ as a function of the initial lateral distance $|\Delta y_i|$ between two pedestrians walking in opposite direction~\cite{corbetta-pre-2018} Error bars indicate a single standard deviation.
    \textbf{(C)} Lateral force field for varying relative coordinates $\Delta x$, $\Delta y$ of a neighbor $j$ walking in opposite direction. Positive values (depicted in red) indicate repulsion, leading to avoidance. 
    \textbf{(D)} Lateral force field as in C, but for a neighboring pedestrian $j$ walking in the same direction. The dashed line indicates the 0-interaction manifold, and the thick black line indicates an ellipsoid fit to this manifold.
    \textbf{(E)} Acceleration induced by interaction with a random crowd of varying size. Crowds consist of individuals with positions randomly selected from a lattice or sampled from a Gaussian distribution. The vertical dashed line indicates $N^* \approx 12$.
    \textbf{(F)} Acceleration resulting from interaction with a crowd with varying spacing. \modelname{} is compared to classical and top-$k$ social force model with and without vision constraint. Only a vision-driven top-$k$ force demonstrates a similar decay as \modelname{}.
     }
    \label{fig:model-interpretation}
\end{figure}

We consider two classes of virtual experiments, sketched in Figure~\ref{fig:model-interpretation}A. First, we consider a diluted regime in which interactions are limited to two pedestrians, and compare our results to a large-scale study on real-world avoidance dynamics~\cite{corbetta-pre-2018}. 
Second, we consider a set of experiments all entailing one pedestrian reacting to a crowd of variable size facing them, and investigate the scaling of $N$-body interactions in crowds. Further details of the experimental scenarios can be found in the Supplemental Material.

\subsection{Diluted interactions}
Similar to~\citet{corbetta-pre-2018}, we consider two pedestrians that walk in opposite direction. To avoid bias of the geometry, the pedestrians are initialized in random position within the domain. Moreover, since \modelname{} is trained to simulate pedestrians moving leftwards, we model the pedestrian moving rightwards with point-symmetric perception and dynamics. In Figure~\ref{fig:model-interpretation}B, we report the average lateral distance when the two pedestrians are closest, $|\Delta y_s |$, as a function of their lateral distance when they first appeared, $|\Delta y_i |$, following~\cite{corbetta-pre-2018}. 

Similar to the findings in~\cite{corbetta-pre-2018}, head-on collisions ($|\Delta y_i |\approx 0$) yield an average offset of $|\Delta y_s |\approx 75\,$cm. Additionally, our results recover the correct, interaction-free, asymptotic behavior ($|\Delta y_s | \approx |\Delta y_i |$) when $|\Delta y_i |$ is large. This marks a difference with respect to the saturating trend in~\cite{corbetta-pre-2018} resulting from the finite size of their considered domain. 
Figure~\ref{fig:model-interpretation}A-1 shows a histogram of lateral distances of pairs walking in opposite directions in the training data, filtered for those cases where there is only a single neighbor within a radius of $r=7.5\,m$. This demonstrates that pairwise avoidance is in-distribution, as a compatible range of values for $|\Delta y|$ are observed in the training data.

As we model accelerations (Eq.~\ref{eq:first-probab-accel}), we can further investigate the avoidance interaction structure. Since the overall acceleration may include a desired velocity effect, analogous to $f$ in social force (Eq.~\ref{eq:social-force-linear}), we distill the mean interaction on a pedestrian $i$ as follows:%
\begin{equation}\label{eq:acc-net-force}
    g(i \mid \mathcal{N}(i)) = \mathbb{E}_{\ddot{x}_i\sim p_\theta, \ddot{x}^\emptyset_i \sim p_\theta}\left[\ddot{x}_i - \ddot{x}^\emptyset_i\right],
\end{equation}
where both $\ddot{x}_i$ and $\ddot{x}^\emptyset_i$ are distributed according to $p_\theta$ in Eq.~\ref{eq:dist_acc_app}, and for $\ddot{x}^\emptyset_i$ we query the model with the same scenario as $\ddot{x}_i$ but without any neighbors ($\mathcal{N}(i) = \emptyset$). Note that the same procedure for a classical social force model (Eq.~\ref{eq:social-force-linear}) would discount the self-propulsion term $f$ and thus measure the cumulative interaction.

Figures~\ref{fig:model-interpretation}C and~\ref{fig:model-interpretation}D show the lateral attraction and repulsion of a pedestrian $i$ relative to a neighbor $j$ as a function of $j$'s relative position $\left(\Delta x, \Delta y\right)$, respectively for $i$ and $j$ walking in parallel and in opposite direction. 
In the opposite case (Figure~\ref{fig:model-interpretation}C), we observe a strong repulsive force which acts at long ranges. Notably, the force vanishes to 0 when $\Delta y \approx 0$, due to neural networks modeling continuous functions and the avoidance force being an odd function, which prevents a nonzero force from emerging at $\Delta y \approx 0$. 
A long-range lateral force with similar structure has also been proposed in~\cite{corbetta-pre-2018}. 
At $20^\circ$, the angle inducing a cone-shaped field-of-view proposed in~\cite{corbetta-pre-2018} is remarkably similar to an angle that aligns well with the contours of our results ($26.6^\circ$), although at shorter ranges, the trumpet-shaped contours deviate from any cone, aligning with a personal space of nonzero width. 

When the neighbor $j$ walks in the same direction (Fig.~\ref{fig:model-interpretation}D), the interaction force includes a short-range repulsion and a long-range attraction. Overall, the interaction force vanishes along a comfort distance manifold, which is compatible with the nearest-neighbor position distribution shown in Figure~\ref{fig:model-validation}D and roughly follows the shape of an ellipse.
Similar to Figure~\ref{fig:model-interpretation}A-1, Figure~\ref{fig:model-interpretation}A-2 shows that a variety of lateral distances can be observed for a single neighbor within $r=7.5\,$m walking in the same direction.

\subsection{N-body interactions}
We focus on the simplest $N$-body interaction case: a pedestrian facing a crowd of static individuals in a random configuration. We consider three randomized crowd configurations, sketched in the right panel of Fig.~\ref{fig:model-interpretation}A: (1) crowds in which the pedestrians are located at random sites of a lattice, with a spacing of $0.5\,$m and a width of $2\,$m; (2) crowds in which the pedestrians' locations are sampled from a Gaussian distribution; and (3) crowds of a fixed size of 18 individuals, organized in a lattice with a total width of $3\,$m, and an additional randomized spacing between the center columns of the lattice. 

Experiments are again repeated across many random positions in the domain, and any acceleration resulting from self-propulsion is discounted as in Eq.~\ref{eq:acc-net-force}. The histogram in Figure~\ref{fig:model-interpretation}A-3 shows the frequency with which a pedestrian has a certain amount of neighbors within a $5\,$m radius in the training data, counting only those neighbors that are in front of the pedestrian. The histogram demonstrates coverage over the range of crowd sizes considered in these experiments in the training data distribution.

Figures~\ref{fig:intro-figure}E and~\ref{fig:model-interpretation}E report the mean acceleration as a function of the crowd size. In all cases, the acceleration saturates as the opposing crowds overcomes a critical size of $N^*\approx 12$ pedestrians. Although a classical social force model scales linearly in the crowd size due to the summation of pairwise interactions (Fig.~\ref{fig:intro-figure}E), we observe that a variant which takes only the top-$k$ ($k=12$) strongest forces into account recovers the saturation accurately, regardless of the randomization strategy, as shown in Fig.~\ref{fig:model-interpretation}E. 
Figure~\ref{fig:model-interpretation}F considers interactions between a pedestrian and crowds with varying spacing between the center columns. Here, both classical and top-$k$ social force models fails to recover the force measured in our experiments. However, Fig.~\ref{fig:model-interpretation}C revealed an anisotropic interaction structure resulting from an avoidance force only within a limited field-of-view of approximately $2 \times 26.6^\circ$. After equipping the top-$k$ social force model with such a field-of-view constraint, the force qualitatively aligns with the measured interactions, dropping rapidly as the path before the pedestrian clears up with increasing spacing in the center of the crowd. 

As a whole, our investigation reveals that the crowd dynamics in the station are compatible with a force that is primarily driven by long-range interactions within a constrained field-of-view. Moreover, these interactions scale in a top-$k$-like fashion and start to saturate around $k \approx 12$ neighbors.

\section{Discussion}\label{sec:conclusion}

Improving our understanding of crowd dynamics has long relied on two paradigms: laboratory studies and real-world measurement campaigns. Although laboratory experiments offer full parametric control, their measurements are limited in statistical resolution. In contrast, real-world measurements allow for virtually unlimited data acquisition, but the lack of parametric control leads to an entanglement of interplaying factors, hindering the isolation of individual effects. Moreover, despite their large scale, the observations provide only sparse coverage of the large space of all possible situations.
Consequently, analyzing real-world data has typically required labor-intensive, tailored procedures to aggregate measurements and to discount external confounding effects. 

Here, we have proposed an approach that combines the strengths of both the laboratory and real-world paradigms. Its backbone is a neural simulator, \modelname{}, trained on a large real-world dataset. \modelname{} represents crowds as dynamic graphs and leverages state of the art developments in generative AI and geometric deep learning to model individual pedestrian movement. Our rigorous validation against key statistical features of crowds confirmed \modelname{}'s ability to perform accurate simulations of crowd dynamics.

We used \modelname{} to investigate interactions in crowds through virtual surrogate experiments, which provide the controllability of lab experiments while maintaining scalability. We first validated this novel approach by examining the prototypical case of pairs of pedestrians in mutual avoidance.  
While not trained specifically on this scenario, we found that \modelname{}'s simulations are in strong agreement with previously reported experimental findings based on thousands of real-world measurements, recovering both short-range and asymptotic avoidance trends. Additionally, we found that interactions are strongest within a narrow field-of-view, in line with a bottom-up model for pairwise avoidance.

We then extended our investigation to more complex and long sought for $N$-body interactions. Our results demonstrate that the interaction strength saturates as the crowd size grows and exceeds $N^* \approx 12$ neighbors, and that a vision-constrained top-$k$ social force model leads to interactions that are in remarkable agreement with this result. 
Overall, our findings emphasize the visual nature of pedestrian interactions and are compatible with a topological interaction structure, as opposed to a distance-based structure. Prior work has shown that such topological interactions are crucial in the dynamics of flocks of birds~\cite{ballerini2008interaction}. In this light, our results provide further evidence for topological interaction in dynamical systems of active agents.

Crowds are an excellent example of real-world dynamics with entangled, heterogeneous physical effects and measurements that are sparse in the set of all possible scenarios.
Collective traffic and animal behavior are other systems of interacting agents that exhibit similar characteristics, and more broadly opinion dynamics, financial markets, and smart energy grids all exhibit stochastic dynamics with complex $N$-body interactions. This work not only demonstrates how neural simulators can be used to uncover the physics behind crowd dynamics, but also establishes their potential as a technology for data-driven scientific discovery in a variety of societally relevant domains.

\clearpage

\section{Methods}\label{sec:methods}

\subsection{Generative model}\label{sec:methods-genmodel}
\begin{figure}[h]
    \centering
    \includegraphics[width=0.5\linewidth]{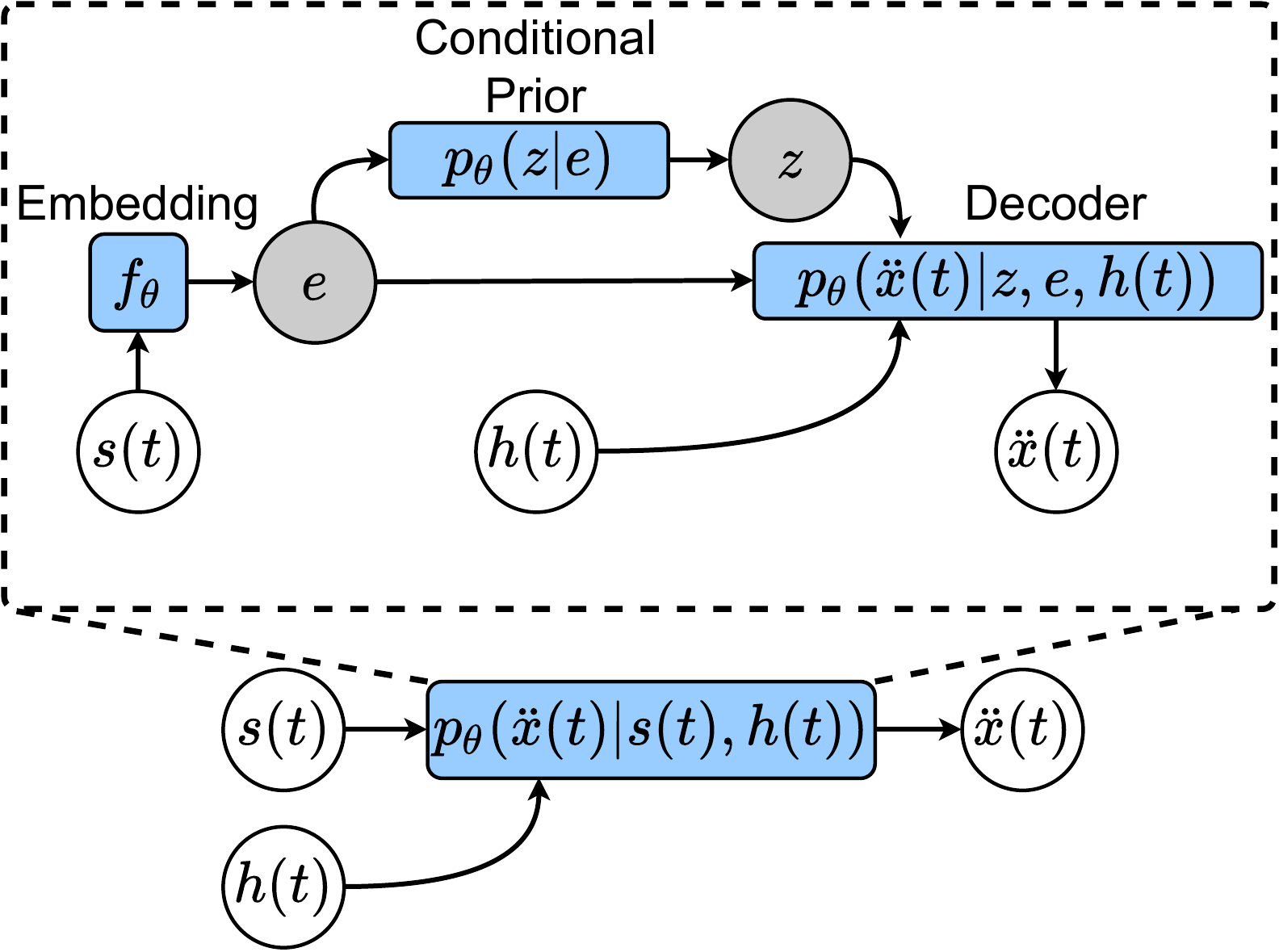}
    \caption{Generative model schematic.}
    \label{fig:model-schematic}
\end{figure}
To model Eq.~\ref{eq:first-probab-accel}, we start from the traditional social force model, in which the distribution over accelerations of each pedestrian $i$ is conditioned on the current positions and velocities of the all pedestrians:
\begin{equation}\label{eq:dist_methods}
p_\theta \left(\ddot{x}_i(t) \mid s(t) \right)= p_\theta \left(\ddot{x}_i(t) \mid x_i(t), \dot{x}_i(t), \left\{x_j(t), \dot{x}_j(t) \mid j \neq i \right\} \right).
\end{equation} 
To derive an informative representation of the current state $s(t)$, we first process it with an embedding network $f_\theta$, such that $e = f_\theta\left(s(t)\right)$. Then, we choose to model Eq.~\ref{eq:dist_methods} with a conditional VAE, following state-of-the-art methods for probabilistic trajectory modeling like trajectron++~\cite{Salzmann2020trajectron++} and EPNS~\cite{minartz2023epns}:
\begin{align}\label{eq:markovian}
    p_\theta \left(\ddot{x}_i(t) \mid e(t) \right) = \int_z \underbrace{p_\theta \left(\ddot{x}_i(t) \mid e(t), z \right)}_\text{decoder} \cdot \underbrace{p_\theta \left(z \mid e(t) \right)}_\text{conditional prior} dz.
\end{align}
Here, both the decoder and the conditional prior are multivariate Gaussian distributions parameterized by the neural network architecture. Although the formulation of Eq.~\ref{eq:markovian} allows for the modeling of expressive distributions, it is incapable of expressing non-Markovian dynamics. Consequently, to extend Equation~\ref{eq:markovian} to account for non-Markovian effects, the VAE's decoder is also conditioned on a representation of all histories $h(t) = \left\{h_j(t)\right\}_{j=1}^N$ via a recurrent unit, following state-of-the-art approaches like~\cite{Salzmann2020trajectron++}. This yields the following generative model:
\begin{align}\label{eq:nonmarkovian-vae}
    p_\theta \left(\ddot{x}_i(t) \mid s(t), h(t) \right) = \int_z p_\theta \left(\ddot{x}_i(t) \mid e(t), h(t), z \right) p_\theta \left(z \mid e(t) \right) dz.
\end{align}
To sample from the distribution specified in Eq.~\ref{eq:nonmarkovian-vae}, in principle we should use ancestral sampling and first draw a realization of the latent variables $z$, and then draw a realization of the accelerations $\ddot{x}_i$ conditioned on $z$. However, it is common practice in VAEs to only sample from $z$ and simply take the mode of $p_\theta(\ddot{x}_i \mid e(t), h(t), z)$. In our case, we observe a trade-off between both strategies. Sampling from the decoder results in better diversity, but lower sample fidelity, while taking the mode of the decoder results in higher quality individual samples, but lower diversity. This exemplifies the fidelity-diversity trade-off commonly encountered in generative models~\cite{xiao2022trilemma}. In order to prioritize diversity or sample quality at inference time, we introduce a \emph{temperature} parameter $T$ with which we multiply the decoder's standard deviation for sampling. $T=T^0=0$ denotes the `frozen' decoder, where we simply take the mode as a sample. We observe that $T^*=T=0.1$ yields a reasonable trade-off between maintaining individual sampling quality, while introducing improved trajectory diversity; higher values of $T$ introduce too much stochasticity in the autoregressive trajectory sampling rollout, leading to unrealistic trajectories.

\subsection{Model architecture and simulation procedure}\label{sec:methods-arch}

\begin{figure}[h]
    \centering
    \includegraphics[width=\linewidth]{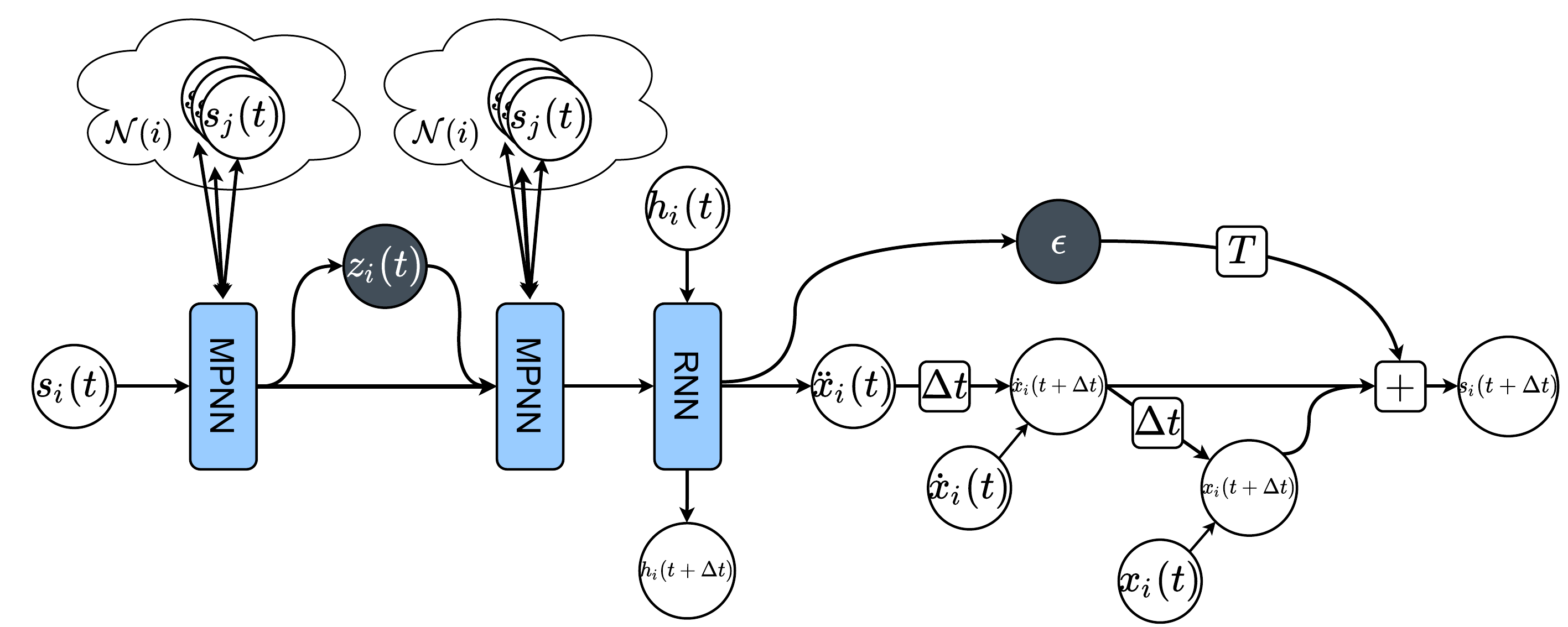}
    \caption{Schematic of model architecture.}
    \label{fig:architecture-schematic}
\end{figure}

Figure~\ref{fig:architecture-schematic} shows a schematic overview of the key architectural components. A message passing GNN serves as backbone architecture. Generally, the $l^\text{th}$ message passing layer is defined as follows~\cite{Gilmer2017mpgnn}:
\begin{align}\label{eq:mpgnn}
    h_i^{l+1} = \psi^l(h_i^l, \bigcup_{j \in \mathcal{N}(i)} \phi^l(h_i^l, h_j^l)), 
\end{align}
where $h_i^l$ is the representation of node $i$ at layer $l$, $\mathcal{N}(i)$ is the set of neighbors of $i$, $\phi^l$ and $\psi^l$ are multi-layer perceptrons, and $\bigcup$ is a permutation-invariant aggregation function. 

Based on the message passing layers defined in Eq.~\ref{eq:mpgnn}, we need to decide how to design $\psi^l$, $\phi^l$, $\bigcup$, and $\mathcal{N}(i)$. 
For simplicity, and similar to existing work~\cite{satorras2021egnn}, we parameterize each $\psi^l$ and $\phi^l$ with two-layer MLPs, where $\psi^l$ models the residual with respect to $h_i^l$, i.e. $h^{l+1}_i = h^l_i + \psi'(\cdot)$. The dimensionality of the hidden representations is fixed at $|h_i^l| = 128$ throughout the model. 
We choose summation as aggregation function $\bigcup$ since it allows for efficiently propagating information about the degree of each node, which relates directly to the local density around a pedestrian. For the neighborhood function $\mathcal{N}(i)$, we use a distance-based thresholding approach to balance efficient propagation of information through the constructed graph with computational efficiency. 
Specifically, at each step, there is an edge between two pedestrians if their distance is less than 5 meters. All GNN modules, depicted in the two leftmost blue rectangles in Figure~\ref{fig:architecture-schematic}, are parameterized with 4 message passing layers, corresponding to the number of iterations of Eq.~\ref{eq:mpgnn}.

To enable the model to take into account non-Markovian effects, we follow state-of-the-art pedestrian trajectory modeling approaches~\cite{Salzmann2020trajectron++} and apply a node-wise LSTM~\cite{Hochreiter1997LSTM} after message passing in the decoder, as shown in the rightmost blue rectangle in Figure~\ref{fig:architecture-schematic}. This recurrent unit takes as input a compressed representation $h_i(t)$ of the history, and outputs an updated node representation as well as an updated history representation $h_i(t + \Delta t)$.

In order to parameterize probability distributions, for example the conditional prior and decoder of the model, node embeddings are mapped to vectors $\mu_i$ and $\sigma_i$ for each node $i$ with 2-layer MLPs. Here, $\mu_i$ and $\sigma_i$ express the mean and standard deviation of the distribution of each dimension -- 64 in the case of the latent space $z$. %
Once the decoder has produced the values of $\mu_i$ and $\sigma_i$ that parameterize its output distribution. These are mapped to a Gaussian distribution over the updated state vector $[x_i(t+\Delta t), \dot{x}_i(t+\Delta t)]$ as follows:
\begin{gather}
\begin{aligned}\label{eq:Euler}
    &\dot{x}_i(t+\Delta t) \sim \mathcal{N}\left(\dot{x_i}(t) + \Delta t \cdot \mu_i^{\ddot{x}}, {\sigma_i^{\dot{x}}}^2\right)\\
    &x_i(t+\Delta t) \sim \mathcal{N}\left(x_i(t) + \Delta t \cdot \mathbb{E}\left[\dot{x}_i(t + \Delta t)\right], {\sigma_i^{{x}}}^2\right), 
\end{aligned}
\end{gather}
where $\mu_i^{\ddot{x}}$, $\sigma_i^{\dot{x}}$ and $\sigma_i^x$ are the outputs of the decoder. Note that the $\sigma_i$ parameterize the velocity and position standard deviation directly, and only the means of the distributions are propagated through the integration scheme following~\cite{minartz2023epns}.

To promote the accurate modeling of social interactions, we introduce an \emph{edge correction step} (not shown in Fig.~\ref{fig:architecture-schematic}), which operates on the positions $x_i$ only, inspired by~\cite{Satorras2021ENNF}. After calculating the tentative positions from Eq.~\ref{eq:Euler}, we take the edge embeddings of the final message passing layer $\phi^\text{out}$ (Eq.~\ref{eq:mpgnn}), which we denote by $h_{ij}$, and pass them through an edge correction MLP $\phi^e: \mathbb{R}^{|h_i|} \rightarrow \mathbb{R}^2$. The edge correction on a pedestrian $i$ is then calculated as follows:
\begin{equation}\label{Eq:edge-correction}
    x'_i = x_i + \sum_{j \in N(i)} \phi^e(h_{ij}) \odot \frac{x_j - x_i}{C + \norm{x_j - x_i}^2}
\end{equation}
Where $x'_i$ are the edge-corrected node positions, $\odot$ denotes element-wise product, and $C$ is a constant, which we set to 1 as in~\cite{Satorras2021ENNF}.

Finally, to facilitate more effective message passing, we experimented with introducing three virtual node types, where in the message passing scheme of Eq.~\ref{eq:mpgnn} we used separate weights for $\psi$ and $\phi$ for the different node types and combinations thereof respectively. The first type is the exit node, which we locate at the exit of the platform (see Fig.~\ref{fig:intro-figure}A) and has outgoing edges to all pedestrians. During preliminary experimentation, we found that the incorporation of such nodes, which allows every pedestrian to `see' where the exit is, improved performance empirically.

The other two node types are designed for encoding the system geometry (the edges of the platform and the convenience store) and enabling more effective long-range communication. To encode the geometry, we placed a mesh of \emph{geometry nodes} on the boundaries of the convenience store and platform, with outgoing edges to pedestrians within $5\,$m from them. To facilitate long-range communication, we put \emph{global nodes} on a $3 \times 6$ evenly spaced grid on the map. The global nodes are connected to their neighboring global nodes, and each pedestrian is connected to her nearest global node. However, we find that both of these additions did not affect performance, and local communication via iterative message passing with edges no longer than $5\,$m is sufficient to achieve the presented results. As such, we discard them in our final model. Nevertheless, we envision that these virtual nodes can be useful for extending our model to generalize over different systems and parameters, for example train stations with different layouts or varying crowd management strategies.

\subsection{Training procedure}\label{sec:methods-train}

To train the model, we make use of the Evidence Lower Bound (ELBO) of the conditional Variational Autoencoder as optimization objective for stochastic gradient descent. Given the model in Eq.~\ref{eq:nonmarkovian-vae} and the mapping from accelerations to velocities and positions as laid out in Eq.~\ref{eq:Euler} and~\ref{Eq:edge-correction}, the ELBO is formulated as follows:
\begin{gather}
\begin{aligned}\label{eq:elbo-AR-vae}
    ELBO(s(t + \Delta t)) &= \underbrace{\mathbb{E}_{z \sim q_\phi\left(z \mid s(t), s(t + \Delta t)\right)} \left[\log\left(p_\theta\left(s(t + \Delta t)  \mid s(t), h(t), z \right) \right)\right]}_\text{reconstruction loss}\\
    &- \underbrace{KL\left[q_\phi\left(z \mid s(t), s(t + \Delta t)\right) \,||\, p_\theta\left(z | s(t)\right)  \right]}_\text{KL regularization of latent space},
\end{aligned}
\end{gather}
where $q_\phi$ is the encoder distribution used during training to approximate the true but intractable posterior $p_\theta(z \mid s(t), s(t + \Delta t))$, also parameterized by a GNN, and $p_\theta\left(s(t + \Delta t)  \mid s(t), h(t), z \right)$ is constructed as explained in Eq.~\ref{eq:Euler} and~\ref{Eq:edge-correction}. However, it is well known that autoregressive latent variable models suffer from latent variable collapse issues during training, as noted in for example~\cite{Kingma2016IAF, Bowman2016,tang2021socialpostcollapse, minartz2023epns}. To mitigate this problem, we modify the ELBO using both the free bits trick~\cite{Kingma2016IAF} and KL annealing~\cite{Bowman2016} as follows:
\begin{gather}
\begin{aligned}\label{eq:elbo-AR-vae-modified}
    ELBO(s(t + \Delta t)) &= \mathbb{E}_{z \sim q_\phi\left(z \mid s(t), s(t + \Delta t)\right)} \left[\log\left(p_\theta\left(s(t + \Delta t)  \mid s(t), h(t), z \right) \right)\right]\\
    &- \beta \cdot \sum_{k=1}^{|Z|} \max \left\{\lambda, KL\left[q_\phi\left(z_k \mid s(t), s(t + \Delta t)\right) || p_\theta\left(z_k | s(t)\right)  \right] \right\},
\end{aligned}
\end{gather}
where $\beta$ is the KL annealing parameter and $\lambda$ is the free bits parameter. However, despite applying these strategies, we could not fully prevent posterior collapse from emerging, resulting in the latent variables not fully capturing the randomness inherent in the dynamics. Nevertheless, the model performs well on our statistical evaluation by simply setting $T=0.1$ in the decoder, as explained in Section~\ref{sec:ml-formulation}. On the one hand, this can be interpreted as a Gaussian transition kernel being sufficiently expressive in our case. On the other hand, as the generative modeling field develops, new strategies for mitigating posterior collapse will emerge, which will be directly applicable to \modelname{} and may enable an even more favorable trade-off between sample diversity and fidelity.

Finally, similar to~\cite{Mohamed2022socialimplicit}, we augment the loss with a relative positioning loss term: for all edges between pedestrians in the graph, we calculate the norm of the difference of the relative positions of the ground-truth and predictions, multiply it by a coefficient $\alpha=10$ to achieve a suitable scale compared to the ELBO, and add this to the total loss. We apply multi-step training, iteratively sampling the predicted acceleration using latent variables sampled from the posterior (explained in more detail in~\cite{minartz2023epns}). We start training with a single rollout step, increasing the rollout length every 50 epochs until epoch 200, and every 25 epochs after. Moreover, we use gradient clipping with a threshold of 1.0 to improve training stability. Finally, we train the model for 340 epochs, where in each epoch we sample a single starting point for multi-step training uniformly at random for trajectory in the training data.

\section{Data and code availability}\label{sec:data-avail}
All pedestrian tracking measurements used for the training, validation, and testing of the model are published with~\cite{pouw2024arxiv}. We are working towards publishing all code for data preprocessing, model training, model validation, and running the experiments.

\section{Acknowledgements}\label{sec:acknowledgments}
This work used the Dutch national e-infrastructure with the support of the SURF Cooperative using grant no. EINF-7724. AC acknowledges the support of a starting grant by the Eindhoven Artificial Intelligence Systems Institute (EAISI) of Eindhoven University of Technology.

\clearpage

\bibliography{sn-bibliography}%

\clearpage

\begin{appendices}

\section{Supplemental material}\label{sec:supplemental}

\subsection{Experimental details}\label{sec:sup-exp-details}

In this section, we explain details behind the quantities of the statistical evaluation and behind the virtual surrogate experiments setup.

\subsubsection{Statistical validation.}
For all results related to the statistical model validation, we generate a 60 second simulation starting from the initial conditions in the test set. For pedestrians that enter the system at some point during the observed 60 seconds, we introduce them in the simulation loop at the same moment with the appropriate initial conditions. For pedestrians that leave the system during the 60 second interval, we remove the pedestrian from the simulation loop at the same moment as in the observations. All quantities are then calculated once for the observed test data, and once for the simulated data, and only for those pedestrians that are part of the flow of pedestrians towards the exit, as per the definition in Section~\ref{sec:problem}.

\vspace{0.125cm}
\noindent\textbf{Velocity PDFs -- Fig.~\ref{fig:intro-figure}C.} %
We calculate the pedestrian velocities over time using finite differencing on the simulated trajectory positions. We then plot the observed and simulated empirical probability distribution functions over the velocities, aggregated over all simulation runs, all snapshots in each run, and all pedestrians in each snapshot.

\vspace{0.125cm}
\noindent\textbf{Fundamental diagram -- Fig.~\ref{fig:intro-figure}D.} %
As the fundamental diagram shows the relationship between velocities and densities, we need to have both the velocity and density for each pedestrian in all snapshots. The velocity is determined in the same way as in Fig.~\ref{fig:intro-figure}C -- see the above paragraph. To calculate the density, we base ourselves on the Voronoi method of~\cite{Jia2022voronoi}, but with the added restriction that each pedestrian's cell is cut off to be within the valid domain, i.e. not inside the convenience store and not outside the platform. Subsequently, calculating the reciprocal of the area of a Voronoi cell gives us the local density of the corresponding pedestrian. We then bin the density values into equi-width bins of width 0.1, and calculate the mean and standard deviation of the corresponding velocities. Note that for the density calculation, pedestrians not part of the flow towards the exit are taken into account to calculate the local densities, but only values of pedestrians in the flow are used for the plot.

\vspace{0.125cm}
\noindent\textbf{Velocity mean and covariance field -- Fig.~\ref{fig:model-validation}A.} %
We divide the observations into a grid of $6 \times 20$ cells, corresponding to the number of vertical equi-width intervals by the number of horizontal equi-width intervals. Then, we only consider those cells which have at least 200 observations in total over all trajectories. Then, we calculate the mean and covariance of the velocity vector observations for each cell. We then visualize the mean velocity vector field as a streamplot, while the covariance matrices are plotted at each cell's location as a confidence ellipse. 

\vspace{0.125cm}
\noindent\textbf{Position PDFs -- Fig.~\ref{fig:model-validation}B.} %
The probability distribution functions over vertical positions are generated by filtering all observations on their horizontal position, grouping them in 20 equi-width bins of width $4\,$m, and plotting the probability distribution function over the vertical coordinate for the three bins as indicated in the plot.

\vspace{0.125cm}
\noindent\textbf{Tortuosity -- Fig.~\ref{fig:model-validation}C.} %
We calculate the tortuosity by dividing the length of the actual path traveled by a pedestrian over the length of the shortest possible path, taking into account the geometry of the system:
\begin{equation}
    \text{Tortuosity} = \frac{\int_{t=t_0}^{t_\text{max}} \norm{\dot{x}(t)} dt}{S}.
\end{equation}
$S$ is the shortest distance calculated using Dijkstra's algorithm at a graph with nodes on the exit, at the corners of the convenience store, as well as at the initial and final position of a pedestrian. Edges with finite weights equal to the distance are placed between every pair of nodes where the edge remains in the walkable domain, i.e. the edge does not cross the convenience store. 

\vspace{0.125cm}
\noindent\textbf{Autocorrelation -- Fig.~\ref{fig:model-validation}C.} %
The autocorrelation is calculated according to the formula in Appendix D of~\cite{corbetta-pre-2017}.

\vspace{0.125cm}
\noindent\textbf{Nearest neighbor relative position distribution -- Fig.~\ref{fig:model-validation}D.} %
The procedure for the nearest neighbor distribution function consists of three steps: first, for each pedestrian in the flow towards the exit (see Section~\ref{sec:problem}), we find the nearest neighbor, which can be a pedestrian that is not in the flow, and find the relative positioning of the neighbor to the pedestrian. Second, we reorient the relative position of the nearest neighbor relative to the direction of the velocity vector of the pedestrian. Third, we apply Gaussian kernel density estimation to the reoriented positions to get the distribution over nearest neighbor positions, using a maximum radius of $1.5\,$m, a $50\times50$ grid, and a bandwidth parameter $\sigma=0.05$ as parameters for the estimation procedure.

\subsubsection{Physical interpretation of the learned dynamics}\label{app:exp-setup-details}

\vspace{0.125cm}
\noindent\textbf{Interaction force scaling -- Fig.~\ref{fig:intro-figure}E.} %
We sample the crowd size uniformly at random between 1 and 25 people, and place the crowd at random locations in a regular lattice spaced $0.5\,$m apart. The width and depth of the lattice is 2.5 meters, corresponding to a lattice with $5 \times 5$ locations. The distance of the pedestrian to the crowd is sampled uniformly at random between $2\,$m and $2.5\,$m, and the pedestrian is positioned exactly facing the center of the lattice. The result is a sampled crowd configuration of size $N$, as well as the relative positioning and velocity of a pedestrian walking towards the middle of the possible locations of the individuals in the crowd. We now place this configuration at a random location to the right of the convenience store and discard those samples where more than 5\% of the individuals are sampled inside the store. We then calculate the net acceleration the pedestrian experiences from the crowd according to the model as in Eq.~\ref{eq:acc-net-force}. 

This procedure is repeated for 10 different values of the velocity with which the pedestrian walks towards the crowd, sampled between $0.7\,$m/s and $1.15\,$m/s. For each velocity value, we repeat the experiment for 300 locations. Finally, for each combination of velocity and location, we sample 10 different crowd configurations. The results are then grouped by crowd size to calculate the mean interaction strength.

For the linear and top-k social force, we use the exact same procedure as described above, but now the accelerations are predicted by a social forcing model. The linear social force model is a simple isotropic social force model with pairwise interaction $g$:
\begin{gather}
\begin{aligned}\label{eq:sf-eq-app}
    \vec{r}_{ij} &= x_i - x_j,\\
    U(\norm{\vec{r}_{ij}}) &= U^0 e^{-\norm{\vec{r}_{ij}} /R},\\
    g(\vec{r}_{ij}) &= - \nabla_{\vec{r}_{ij}} U(\norm{\vec{r}_{ij}}).
\end{aligned}
\end{gather}
The top-k variant sums only the top-k strongest pairwise interactions to get the total social force on the pedestrian, and discards the others. The parameters of both the linear and top-k model are $U^0=0.3\,$m$^2$/s$^2$ and $R=2\,$m, which were chosen such that the linear social force model's slope approximately aligns with \modelname{} in the small-N regime.

\vspace{0.125cm}
\noindent\textbf{Counterflow transfer function -- Fig.~\ref{fig:model-interpretation}B.} %
The pairwise avoidance experiment is set up to be comparable to the procedure behind the empirical results of~\citet{corbetta-pre-2018}. First, we initialize the pedestrian and the neighbor with a longitudinal distance of $8\,$m, and sample the initial lateral displacement $\Delta y_i$ of the neighbor uniformly at random between $-2\,$m and $2\,$m. The velocity of both the pedestrian and the neighbor is initialized at $v=0.5\,$m/s, with opposing horizontal directions. We then start a simulation using \modelname{} for 11 seconds. Note that, since the model is trained to model the flow of pedestrians towards the exit, we assume point-symmetric accelerations for the neighbor moving in the opposite direction. After the simulation, we find the time $t_\text{min}$ when the horizontal distance between the two pedestrians is minimal, and measure the lateral distance $|\Delta y_s|$ at the time of passing $t_\text{min}$. This procedure is repeated for 100 different locations on the map, for 300 repetitions per location.

\vspace{0.125cm}
\noindent\textbf{Pairwise avoidance force field -- Fig.~\ref{fig:model-interpretation}C.} %
To measure the lateral forcing corresponding to avoidance behavior, we initialize a pedestrian with a velocity of $v=1\,$m/s walking leftwards at random locations on the map. We then randomly sample a neighbor with relative displacement $\Delta x \in \left[-3\,\text{m}, 0\,\text{m}\right]$, $\Delta y \in \left[-3\,\text{m}, 3\,\text{m}\right]$, walking with the same velocity in opposite direction. This procedure is repeated for 300 locations evenly spaced on the map, with 100 repetitions per location. 

We calculate the acceleration as in Eq.~\ref{eq:acc-net-force} on the pedestrian, multiply it by two to reflect the point-symmetry assumption, and take the lateral component, denoted as $f_y$, to reflect avoidance forcing. We then multiply $f_y$ with the sign of $\Delta y$ such that the sign of the resulting quantity is positive if the forcing is repulsive, and negative if it is attractive.

\vspace{0.125cm}
\noindent\textbf{Parallel walking force field -- Fig.~\ref{fig:model-interpretation}D.} %
In the case of the neighbor walking in parallel direction, we initialize a pedestrian and neighbor with a velocity of $v=1.34\,$m/s walking leftwards at random locations. The relative displacement of the neighbor $\left(\Delta x, \Delta y\right)$ is randomly sampled from $\left[-3\,\text{m}, 3\,\text{m}\right]^2$. This procedure is repeated for 300 locations evenly spaced on the map, with 50 repetitions per location. 

After calculating the average acceleration as in Eq.~\ref{eq:acc-net-force} for a neighbor located at a specific $\left(\Delta x, \Delta y\right)$, we subtract the acceleration of a neighbor located at $\left(-\Delta x, -\Delta y\right)$ to reflect the relative acceleration under the assumption of point-symmetric dynamics. We again take the lateral component of the interaction $f_y$ and multiply it with the sign of $\Delta y$ such that positive quantities reflect repulsion and negative quantities reflect attraction.

\vspace{0.125cm}
\noindent\textbf{N-body scaling with different crowd configurations -- Fig.~\ref{fig:model-interpretation}E.} %
The lattice-based crowd configurations are taken as in Figure~\ref{fig:intro-figure}E. For the Gaussian crowd configuration, we sample the relative coordinates of each individual following a Gaussian distribution, with $\Delta x \sim \mathcal{N}\left(\mu_{\Delta x}, 0.5^2\right)$, $\Delta y \sim \mathcal{N}\left(0, 0.5^2\right)$,
where $\mu_{\Delta x} \sim \mathcal{U}\left[-3.5, 2.0\right]$ and all units are in meters. The parameters of the comparable top-$k$ social force model are $U_0 = 0.21$m$^2$/s$^2$ (compared to $0.3$m$^2$/s$^2$ in the lattice case) and $R=2.0\,$m (identical to the lattice case).

\vspace{0.125cm}
\noindent\textbf{Anisotropic N-body interactions -- Fig.~\ref{fig:model-interpretation}F.} %
We start from the lattice-based crowd configuration as explained in figure~\ref{fig:intro-figure}E, with the following changes: (1) to make sure all individuals in the crowd remain within the $5\,$m edge cutoff radius, we increase the lattice width to $3\,$m so that we require less depth, and (2) the distance to between the pedestrian and the crowd is always $2.5\,$m. We then construct a crowd of 18 static individuals, organized in a $6\times 3$ lattice with $0.5\,$m spacing. We then add a random additional spacing $d_\text{gap}$ between the center two rows of the lattice, which is right in front of the pedestrian. $d_\text{gap}$ is sampled uniformly between $0\,$m and $4\,$m. For the field-of-view social forcing models, only forces of individuals in the crowd that are within $2 \times 26.6^\circ$ are taken into account -- all other social forcing parameters are as described in Figure~\ref{fig:intro-figure}E.

\subsection{Collisions}\label{sec:sup-add-results}

\begin{figure}[t]
    \centering
    \includegraphics[width=\linewidth]{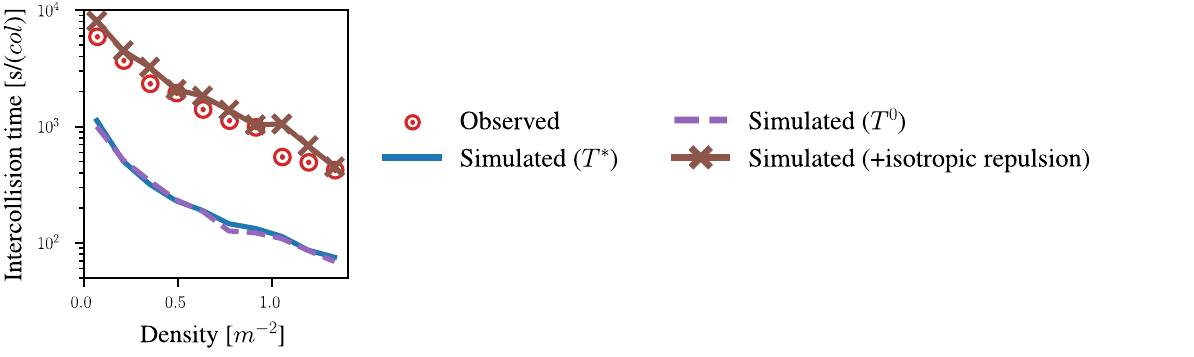}
    \caption{Average intercollision time per pedestrian as a function of density, plotted in log scale. In addition to the $T^*$ model, we show results of a hybrid variant of~\modelname~and social forcing. Here, we exploit domain knowledge from the physics-based modeling field by adding an isotropic social force to the predicted accelerations.} 
    \label{fig:collisions}
\end{figure}

We also study the amount of collisions in the observations and simulations. To this end, we assume each pedestrian to be a circle with a $r=0.2\,$m radius, and consequently define a collision event as the distance between two pedestrians being smaller than $0.4\,$m. Figure~\ref{fig:collisions} shows the intercollision time, defined as the average time between a pedestrian experiencing subsequent collision events, as a function of density. We observe that the intercollision time in the measurements decays exponentially from roughly $10^4\,$s to $10^2\,$s as the density increases from $0\,$m$^{-2}$ to $2\,$m$^{-2}$. In contrast,~\modelname{} exhibits substantially smaller intercollision times for all densities, indicating more frequent collisions than in real life. This observation is in line with earlier observations on ML techniques for pedestrian dynamics simulation~\cite{Jiang_2022_NSP}. We find that this is explained by a lack of sufficient training data of pedestrians at near-collision ranges, as these distances are not socially acceptable. This implies that the model has no chance to learn a strong repulsive forcing at very close ranges directly from data. Consequently, if the model erroneously predicts two pedestrians getting too close to each other, it has not learned to correct this mistake in the next time step.

Interestingly, we found that adding a simple isotropic social force ($U^0 = 0.5\, \text{m/s}^2,\, R=0.125\,$m, see Eq.~\ref{eq:sf-eq-app}) to the predicted accelerations results in an intercollision time that is well-aligned with the measurements, as shown in Fig.~\ref{fig:collisions}. Although not within the scope of this work, this suggests that a hybrid, physics-informed approach is a promising avenue towards even more accurate data-driven simulations.

\end{appendices}

\end{document}